\documentclass[twocolumn,english,groupedaddress, superscriptaddress,aps,pre,10pt]{revtex4-1}
\usepackage[T1]{fontenc}
\usepackage[utf8]{inputenc}
\usepackage{gensymb}
\usepackage{textcomp}
\usepackage{amsmath}
\usepackage{graphicx}
\usepackage{babel}
\usepackage{url}
\usepackage[normalem]{ulem}

\usepackage{color}
\definecolor{blue}{rgb}{0,0,1}

\definecolor{dgreen}{rgb}{0,0.5,0}

\definecolor{dred}{rgb}{0.5,0,0}

\definecolor{purple}{rgb}{1,0.2,1}

\definecolor{dyellow}{rgb}{0.75,0.75,0}

\definecolor{lightBlue}{rgb}{0.,0.5,0.5}

\definecolor{fullRed}{rgb}{1.,0,0}

\newcommand{\revise}{\textcolor{black}}
\begin{document}

\title{Wind Power Persistence is Governed by Superstatistics}
\title{Wind Power Persistence Characterized by Superstatistics}

\author{Juliane Weber}
\affiliation{Forschungszentrum J\"ulich, Institute for Energy and Climate Research
- Systems Analysis and Technology Evaluation (IEK-STE), 52428 J\"ulich,
Germany}
\affiliation{Institute for Theoretical Physics, University of Cologne, 50937 K\"oln,
Germany}

\author{Mark Reyers}
\affiliation{Institute for Geophysics and Meteorology, University of Cologne, Cologne, Germany}

\author{Christian Beck}
\affiliation{Queen Mary University of London, School of Mathematical Sciences,
Mile End Road, London E1 4NS, UK}

\author{Marc Timme}
\affiliation{Chair for Network Dynamics, Center for Advancing Electronics Dresden
(cfaed) and Institute for Theoretical Physics, Technical University
of Dresden, 01062 Dresden, Germany}

\author{Joaquim G. Pinto}
\affiliation{Institute of Meteorology and Climate Research, Karlsruhe Institute of Technology, Karlsruhe, Germany}

\author{Dirk Witthaut}
\email{d.witthaut@fz-juelich.de}
\affiliation{Forschungszentrum J\"ulich, Institute for Energy and Climate Research
- Systems Analysis and Technology Evaluation (IEK-STE), 52428 J\"ulich,
Germany}
\affiliation{Institute for Theoretical Physics, University of Cologne, 50937 K\"oln,
Germany}

\author{Benjamin Sch\"afer}
\email{b.schaefer@qmul.ac.uk}
\affiliation{Queen Mary University of London, School of Mathematical Sciences,
Mile End Road, London E1 4NS, UK}
\affiliation{Chair for Network Dynamics, Center for Advancing Electronics Dresden
(cfaed) and Institute for Theoretical Physics, Technical University
of Dresden, 01062 Dresden, Germany}

\begin{abstract}
Mitigating climate change demands a transition towards renewable electricity generation, with wind power being a particularly promising technology. Long periods either of high or of low wind therefore essentially define the necessary amount of storage to balance the power system. While the general statistics of wind velocities have been studied extensively, persistence (waiting) time statistics of wind is far from well understood. Here, we investigate the statistics of both high- and low-wind persistence. We find heavy tails and explain them as a superposition of different wind conditions, requiring $q$-exponential distributions instead of exponential distributions. Persistent wind conditions are not necessarily caused by stationary atmospheric circulation patterns nor by recurring individual weather types but may emerge as a combination of multiple weather types and circulation patterns. \revise{This also leads to Fr\'echet instead of Gumbel extreme value statistics.} Understanding wind persistence statistically and synoptically may help to ensure a reliable and economically feasible future energy system, which uses a high share of wind generation.
\end{abstract}

\maketitle


\section*{Introduction}
The $2 \degree \text{C}$ target of the Paris agreement \cite{Paris2015} requires a rapid decarbonization of the
energy sector \cite{Rogelj2016,Figueres2017}. The most promising technologies to reach this goal are wind and solar power generation, which have shown a remarkable development in the last decade  \cite{Edenhofer2011,Creutzig2017,IRENA2018,Gotzens2018},
paving the way to a fully renewable energy supply \cite{Jacobson2011,Rodriguez2015}.
However, integrating the specifically important wind power generators \cite{Rodriguez2015} into the power system comes with a large challenge: Wind power generation is  strongly modulated by weather conditions and thus strongly fluctuates on time scales from seconds to weeks or even decades \cite{Milan2013,Olauson2016,Ren2018,Wohland2018}.

A variety of technical measures is currently being developed to cope with these fluctuations in the power system. Virtual inertia \cite{Morren2006}, batteries \cite{Divya2009,Soni2013,Janoschka2015}, or smart grid applications \cite{Fang2012,Schaefer2015} might balance the grid for seconds, minutes or a few hours. For time periods of many minutes or several hours, pumped hydro storage is capable of providing back-up power \cite{Rehman2015}.
However, it remains unclear how to act when low wind conditions persist for several days or weeks.

Long periods characterized by a persistent and quasi-stationary blocking high pressure weather system (which may endure several weeks) lead to sustained low-wind velocities and thus constitute extreme weather events \cite{Meehl2000}, posing a substantial challenge to the operation of highly renewable power systems.  During these periods, the power demand must be entirely satisfied by other renewable generators, backup power plants \cite{Elsner2016} or long-term electricity storage, which is not yet available at that scale \cite{Dunn2011}. Not the average power output of wind farms, but the extreme event statistics is essential when dimensioning the necessary backup options \cite{Huber2014,Rodriguez2014,Rodriguez2015,Schlachtberger2016}. It is assumed that these extreme events without renewable generation occur rarely, but a clear quantitative understanding is missing.

In addition, periods with continuously high-wind power generation have also striking impacts on electricity grids and markets.  A high-wind power feed-in already led to negative electricity prices \cite{Paraschiv2014} and lead to transmission grid congestion \cite{Pesch2014,Wohland2017}. In future highly renewable energy systems, these high-wind periods determine the potential for new applications such as Power-to-Heat or Power-to-Gas \cite{Sternberg2015} or the occurrence of surplus electricity and the need of curtailment \cite{Georgilakis2008,Burke2011,Barnhart2013}.  Again, the question arises: How long can these periods last and how likely do they occur?

To answer these questions, we need to investigate and understand the statistics of long periods with very low or very high power generation by wind  \cite{Elsner2016}. 
While the statistics of wind velocities \cite{Justus1978,Weber2017}, its increment statistics \cite{Boettcher2003,Morales2012,Anvari2016} and the associated power generation \cite{Weber2017a,Wohland2018} have been explored extensively, the persistence of wind \cite{Simiu1996,Koscielny1998} and its extreme event statistics \cite{Nicolosi2010} are less studied and far from well understood.

In this article, we investigate the persistence (waiting time) statistics  of low- and high-wind situations in Europe. 
We thus analyze the duration of periods where wind velocities $v$ constantly stay below or above a certain limiting value.
The study is carried out for various locations in Europe and complemented with an analysis of aggregated power generation for individual countries and a detailed synoptic analysis. We mainly focus on the statistical analysis of the wind data and its interconnection with the synoptic system. Overall, we demonstrate how non-standard statistics are necessary to describe waiting time persistence distributions. Further, we argue that dynamical large-scale weather conditions \cite{Grams2017} contribute to local persistence statistics. This might impact future energy modelling by requiring additional storage capacity.

\section*{Results}

\begin{figure*}
\begin{centering}
\includegraphics[width=1.99\columnwidth]{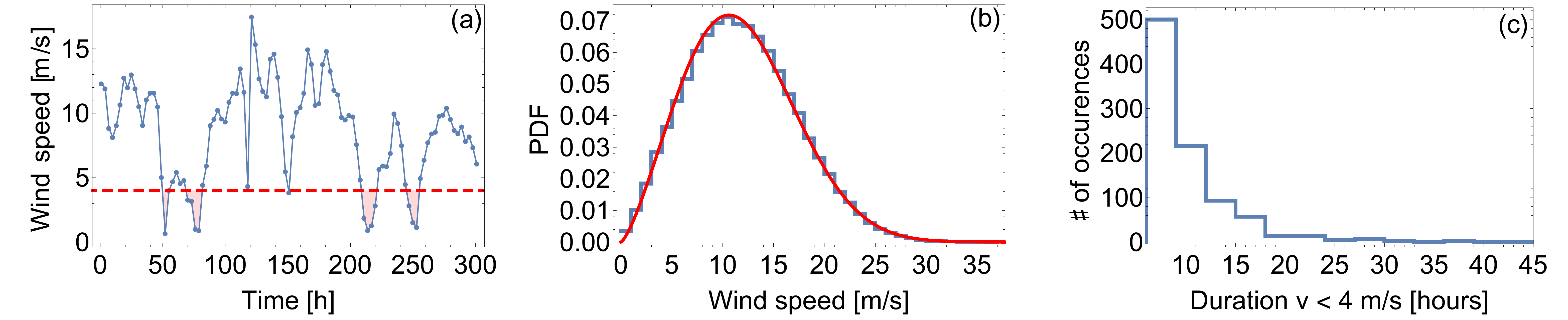}
\par\end{centering}
\caption{\textbf{Extracting wind persistence statistics from trajectory data.}
a: The downscaled ERA-Interim data at Alpha Ventus \cite{Dee2011} provide a trajectory of wind velocities with a 3-hour resolution. 
b: The aggregated wind velocities approximately follow  a Weibull distribution. 
The blue curve reports the recorded data and the red curve depicts the most-likely Weibull distribution, with shape parameter $\alpha\approx 2.36$ and scale parameter $\beta\approx 9.66$.
c: If the wind velocity drops below a threshold of $v=4~\text{m/s}$ (dashed red line in panel~a), we count the full period until it crosses the threshold again as low-wind duration and gather these events for our persistence statistics. 
For high-wind speeds, we analogously employ an upper threshold of $v=12~\text{m/s}$ (not shown). Note that the plots and thresholds use the velocity scaled up from $10$~m to a typical hub height of $100$~m, using a power law, see Methods for details. While the mean velocity in panel (b) is close to the upper threshold, we note that here we are using data from an offshore wind farm location, with typically high wind velocities. Most wind turbines reach their rated power at  $v=12 m/s$ \cite{Ackermann2005} so that higher velocities still lead to the maximum power output.
\label{fig:VelocityStatisticsAndIllustration}
}
\end{figure*}

\subsection*{Wind persistence statistics}
\label{sec:statistics}

Extreme wind conditions represent a major challenge for the operation of future highly renewable power systems. The aggregated wind velocity statistics follow a well-known Weibull distribution \cite{Justus1978,Seguro2000}, which can be used to derive the probability for situations with low and high-wind power generation, see Fig.~\ref{fig:VelocityStatisticsAndIllustration}a.
In contrast, much less attention has been paid towards the temporal patterns of wind. Especially the probability of long durations with low-wind power are of central importance to assess the reliability of renewable power systems and to plan necessary backup infrastructures \cite{Elsner2016,Rodriguez2015,Weber2017}.

Here, we analyze the persistence statistics of wind velocities and wind power using publicly available  wind data sets provided by the EURO-CORDEX consortium \cite{Jacob2014} with high temporal and spatial resolution. 
In particular, we use wind speeds from the ERA-Interim Reanalysis data set \cite{Dee2011} which is downscaled to a high spatial and temporal resolution using the regional model RCA4 \cite{Samuelsson2011}. The wind velocity time series covers a grid all over Europe for a time frame of 31 years from 1980-2010 with 3-hour time resolution. 
The simulations have a horizontal resolution of 0.11\textdegree, such that local orographic effects and the impact of large-scale atmospheric dynamics are captured realistically. ERA-Interim Reanalysis are widely used as boundary conditions for EURO-CORDEX regional climate model simulations, also for wind energy applications, see e.g. \cite{Tobin2016,Moemken2018}. We therefore conclude that this data set forms a reliable basis to identify periods of potentially high and low wind speeds associated with strong and weak synoptic-scale pressure gradients, respectively.

We identify continuous intervals of the wind time series where velocities are below $v<4~\text{m/s}$ because most wind turbines start generating power at this wind speed \cite{Ackermann2005} and classify them as periods with low wind, see Fig.~\ref{fig:VelocityStatisticsAndIllustration} panel~a for an illustration of the procedure and a comparison between aggregated (panel~b) and persistence statistics (panel~c).
Analogously, we record the duration of intervals of constant high-wind velocities $v\geq12~\text{m/s}$ 
as a typical value of the rated wind speed \cite{Ackermann2005}.  While some locations, such as Alpha Ventus (Fig. \ref{fig:VelocityStatisticsAndIllustration}b) have a high average wind velocity, the chosen thresholds are based on the rated power of typical wind turbines \cite{Ackermann2005}.
 Since locations with an abundance of wind return a small number of persistent events with $v<4~\text{m/s}$, we mainly analyze low-wind speed statistics for low-wind locations (e.g. continental regions) and high-wind speed statistics for high-wind locations (e.g. offshore wind farms). Complementary analysis is shown in Supplementary Note 2. Altering the time resolution or introducing a maximum cut-off wind speed has little influence on the statistics (Supplementary Note 6).

Intuitively, persistence statistics should follow an exponential distribution. It arises naturally if the events that cross the threshold, e.g. of $v<4 ~\text{m/s}$, follow a Poisson process \cite{Faris2001,Ross2014,Krause2017}. In this case, the statistics of the waiting time or persistence $d$ are described by the probability density function 
\begin{equation}
   p(d|\lambda_e)=\lambda_e\exp(-\lambda_e d),
\end{equation}
for a fixed exponential decay constant $\lambda_e$, which may assume different values as we discuss below.

When analyzing persistence statistics, the tails of the distribution are of special interest, because they determine the likelihood of extreme events. We use the kurtosis $\kappa$, the normalized 4th moment of the distribution as a measure of how \revise{heavy-tailed the data are \cite{westfall2014kurtosis}, see Methods for a formal definition of the kurtosis}. An exponential distribution has a kurtosis of $\kappa_{\rm exp}=9$ such that a larger value $\kappa>9$ indicates heavy tails.

\begin{figure*}
\begin{centering}
\includegraphics[width=1.8\columnwidth]{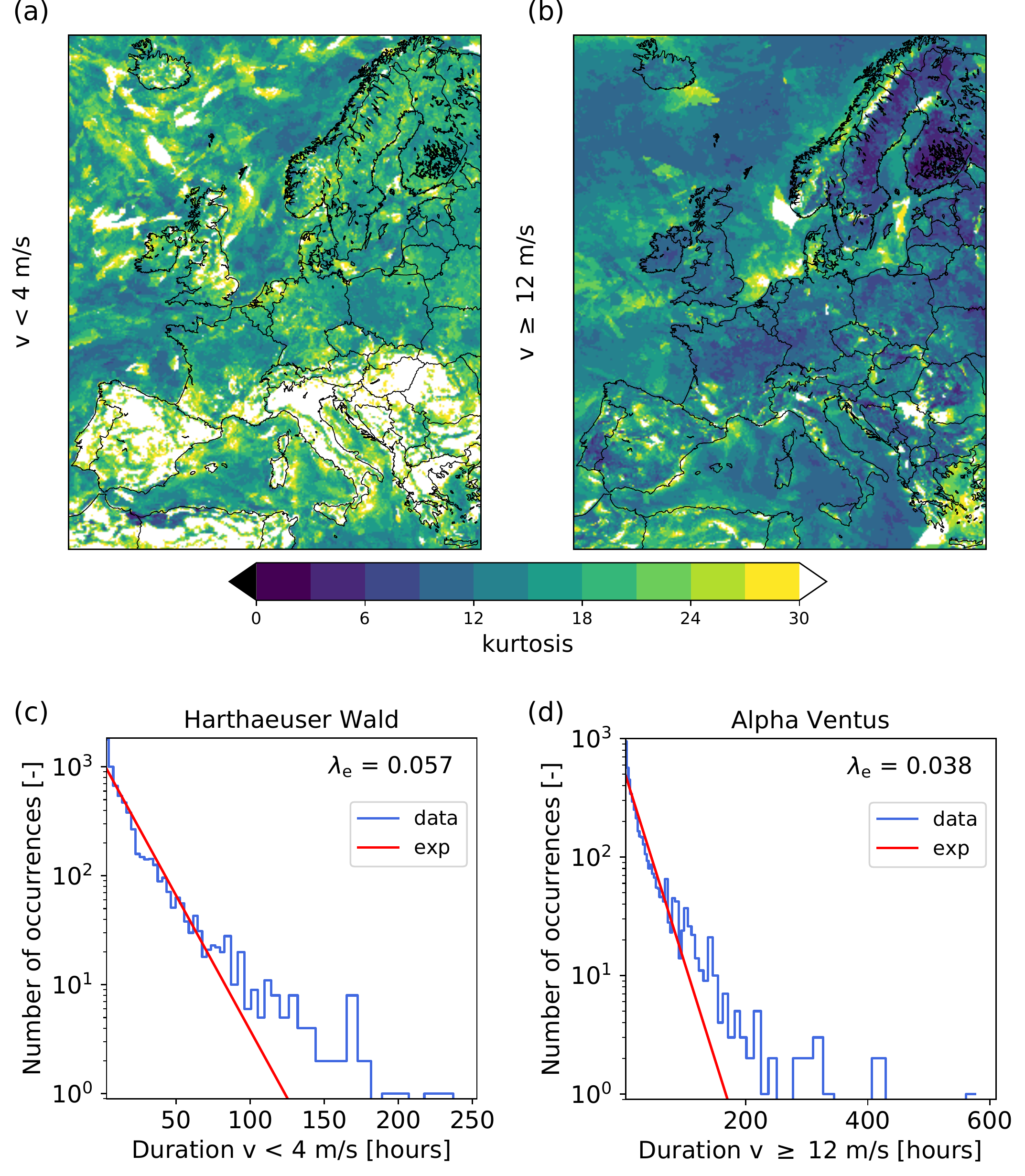}
\par\end{centering}
\caption{\textbf{European wind persistence statistics are heavy-tailed for low- and high-wind velocities.}
The kurtosis of the persistence statistics for low-wind (a) and high-wind (b) states is shown. \revise{A kurtosis of 30 or greater is depicted as white.} If the data were following an exponential distribution, the kurtosis should be $\kappa=9$ so that a kurtosis above this value indicates heavy tails. We show the persistence statistics for two specific locations:
c: Harthaeuser Wald is analysed for low-wind velocities $v<4~\text{m/s}$, while d: Alpha Ventus is used for high-wind velocity analysis $v\geq12~\text{m/s}$.
All analysis is based on the downscaled ERA-Interim data from 1980-2010 \cite{Dee2011}. The blue curves give the data and the red curves depict the most-likely exponential fits. In both cases, the exponential fit underestimates the tails of the distribution, which crucially determine the extreme event statistics.
 Maps were created using Python 2.7.12: \protect\url{https://www.python.org/}.
}
\label{fig:Era-InterimKurtosis}
\end{figure*}


Do wind persistence statistics follow a simple exponential distribution or do they display heavy tails? 
Indeed, analyzing the downscaled ERA-Interim data reveals heavy tails, i.e. many locations in Europe display a kurtosis much larger than 9, see Fig.~\ref{fig:Era-InterimKurtosis}. The strongest heavy tails in terms of kurtosis are observed for the statistics of low-wind states in the countries around the Mediterranean sea. In particular, this includes most parts of the Iberian Peninsula, Southern France, Italy, large parts of the Balkan, Greece and parts of Northern Africa.

Investigating individual locations, we find that the persistence statistics of low or high-wind is not well-approximated by exponential distributions, see 
Fig.~\ref{fig:Era-InterimKurtosis} (c)-(d). 
A maximum likelihood estimate for an exponential distribution at a representative on-shore location far from the coast (the wind farm Harthaeuser Wald, German: "Harthäuser Wald", in South-Western Germany)
strongly underestimates the likelihood of very long durations.
Similarly, the likelihood of very long high-wind situations is underestimated for a typical off-shore location (the wind farm Alpha Ventus in the North Sea).
Interestingly, the low-wind periods at Harthaeuser Wald are much shorter than the high-wind periods at Alpha Ventus. We generally observe this trend of high-wind periods persisting longer than low-wind periods also at other locations, see Supplementary Note 2. Further analysis of different locations in Europe including a map indicating their position is given in Supplementary Note 1. The pronounced tails can be interpreted as a consequence of long-range correlations in the time series, leading to high wind states being followed by further high wind states, see also \cite{Koscielny1998,Anvari2013} and Supplementary Note 7 for a correlation analysis.

We conclude that a refined statistical analysis is necessary to capture the tails of the persistence statistics.

\begin{figure*}
\begin{centering}
\includegraphics[width=1.9\columnwidth]{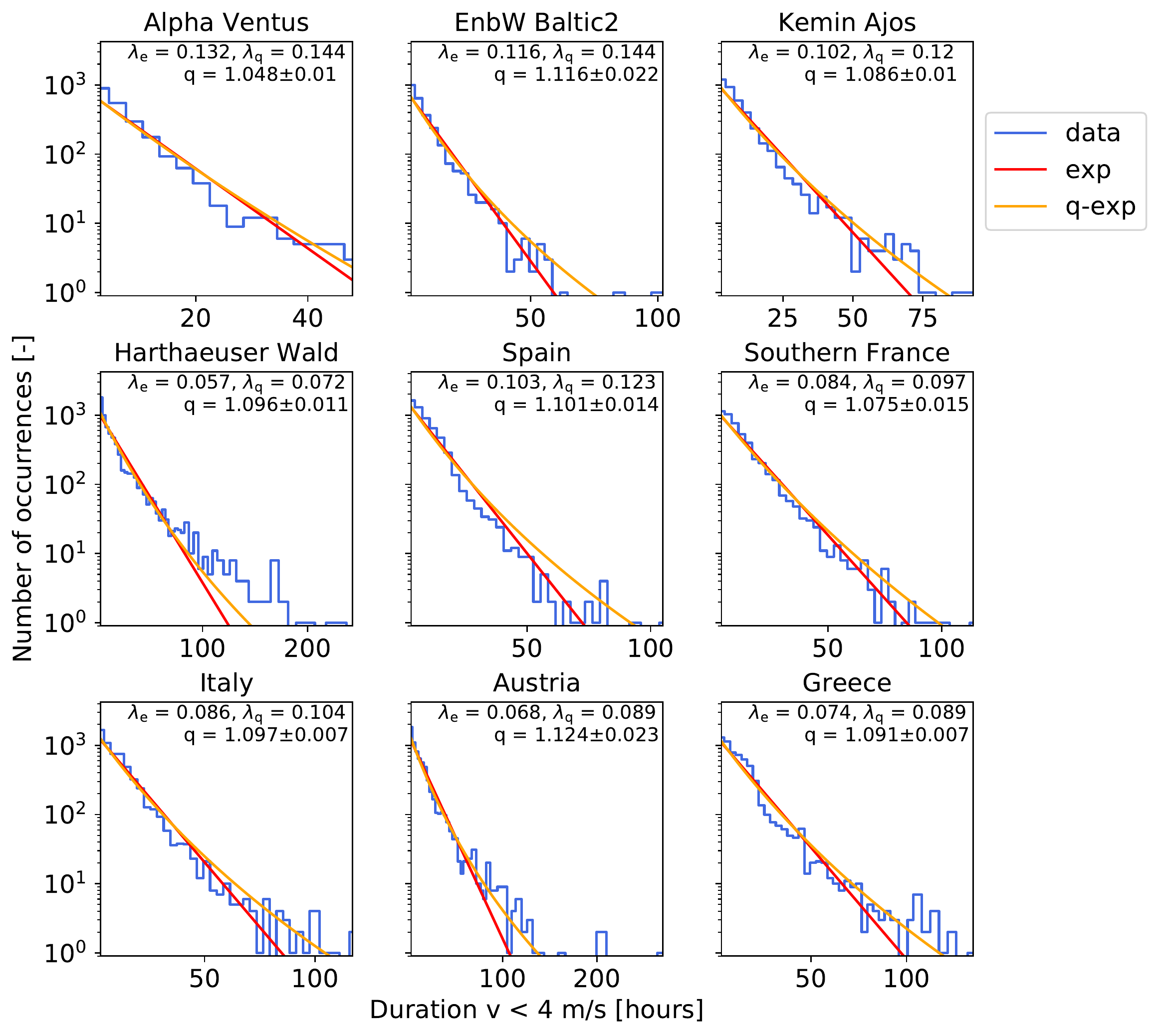}
\par\end{centering}
\caption{\textbf{Distributions are not strictly exponential but better described by $q$-exponentials for low-wind.}
Wind persistence statistics (blue) is shown with the most-likely exponential (red) and $q$-exponential distributions (orange) for 9 selected locations, based on the downscaled ERA-Interim data \cite{Dee2011}. The $q$-values are determined by using the kurtosis of the data, see eq.~(10) in Methods.  Note that the maximum $q$-value derived this way is $q_\text{max}=1.2$. We report the uncertainty of $q$ as a single standard deviation, determined via bootstrapping, see Methods. See also Supplementary Note 1 for a map of the locations.
}
\label{fig:ExpAndQExpFits_lowWind}
\end{figure*}


\begin{figure*}
\begin{centering}
\includegraphics[width=1.8\columnwidth]{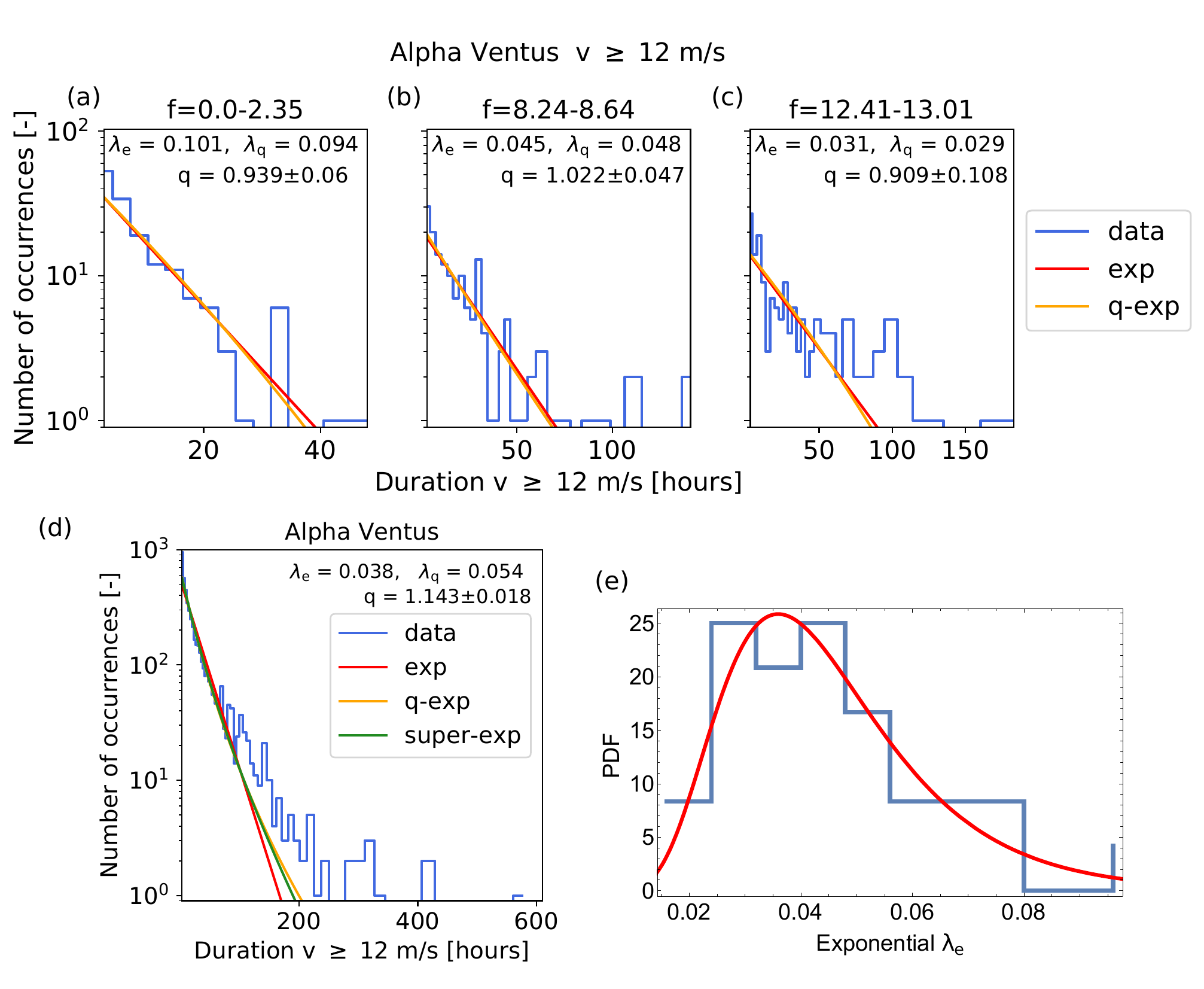}
\par\end{centering}
\caption{\textbf{Persistence statistics approximately follows exponentials for homogeneous pressure.}
High-wind velocity statistics $v>12~\text{m/s}$ are analyzed for Alpha Ventus, based on the downscaled ERA-Interim data \cite{Dee2011}, conditioning the statistics on small bins of homogeneous $f$-parameters (in units of hPa per $1000$~km).
(a)-(c) Plotting both the most-likely exponential and $q$-exponential distributions for small, conditioned subsets, we notice that the $q$-exponential distributions are very close to the exponential ones. The $q$-value is determined by using the kurtosis of the data, see Eq \eqref{eq:QKurtosis}. Note that the maximum $q$-value derived this way is $q_\text{max}=1.2$. On average, the $q$-value is closer to $1$ than in the unconditioned Fig.~\ref{fig:ExpAndQExpFits_lowWind}.
(d) Combining the independent exponential distributions into one super-exponential approximates the $q$-exponential distribution.
(e) The histogram of the individual $\lambda_e$ parameters is approximated by a log-normal distribution, a typical distribution often seen in superstatistics.
We report the uncertainty of $q$ as a single standard deviation, determined via bootstrapping, see Methods.
See also Supplementary Note 4 for detailed discussion and analysis of Harthaeuser Wald.}
\label{fig:IndividualExpFits}
\end{figure*}

\begin{figure}
\begin{centering}
\includegraphics[width=0.9\columnwidth]
{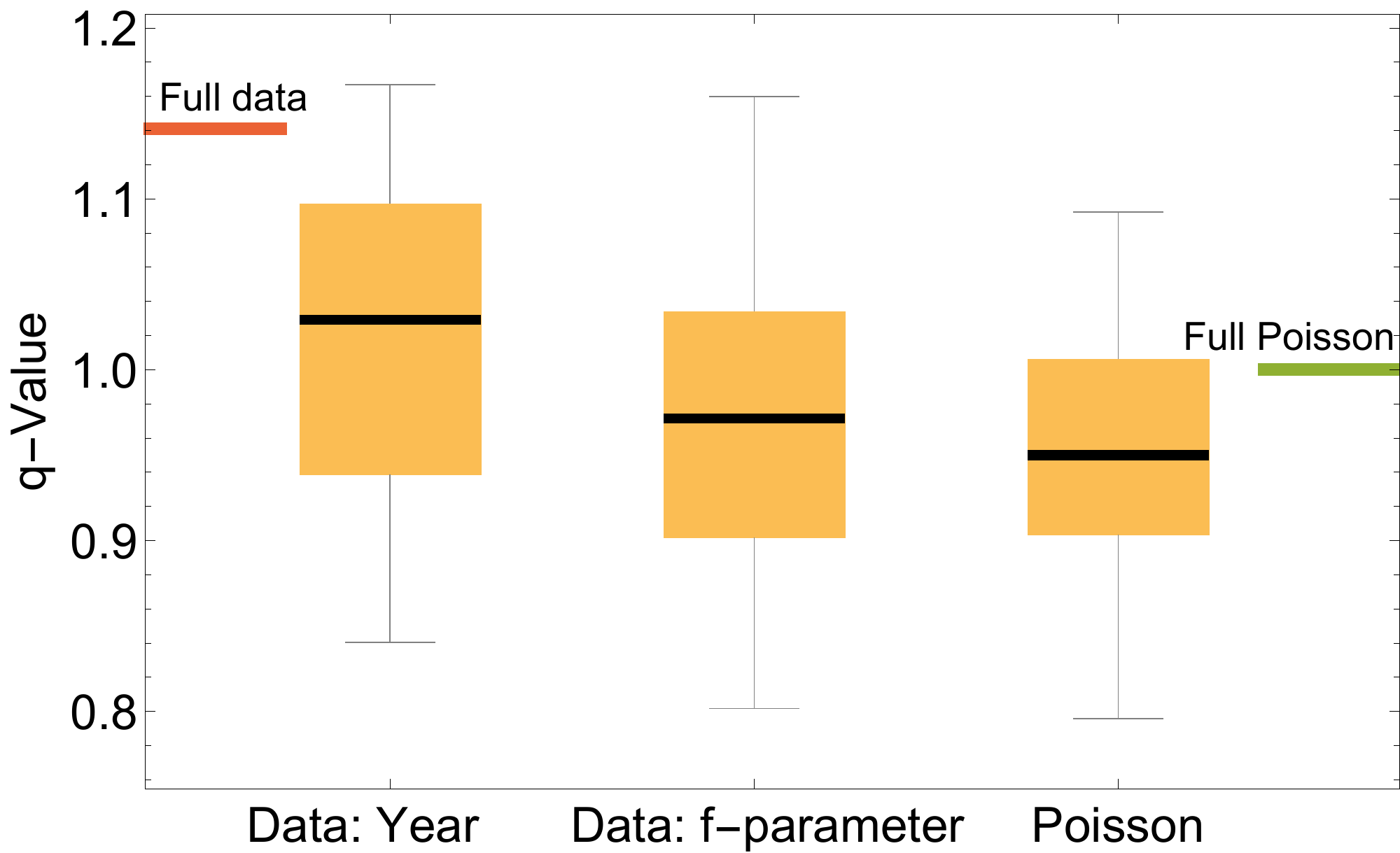}
\par\end{centering}
\caption{
\textbf{Data subsets with homogeneous pressure approximate Poissonian statistics.}
High-wind velocity persistence statistics, $v\geq12~\text{m/s}$, is analyzed at Alpha Ventus, based on the downscaled ERA-Interim data from 1980-2010 \cite{Dee2011}. 
Three different data sets are compared: First, the original data, consisting of 31 years of measurements is split into 31 equally sized data sets, based on the year it was recorded (Data: Year). Alternatively, the data is split  based on approximately homogeneous $f$-parameter (Data: $f$-parameter). Finally, this is compared to an artificial Poissonian process with return times as estimated from the exponential distribution, generating an equal number of data points (Poisson). 
The $q$-values of the full sets are indicated by colored lines at the sides, both for the real data as well as the Poissonian process. 
The full data $q$-value is larger than the $q$-values of most subsets. Furthermore, splitting the data arbitrarily according to calendar years leads to more values at large $q$ than if the data is conditioned on the $f$-parameter. 
Conditioning on the $f$-parameter approximates the Poisson distribution much better than yearly conditioning, when computing the Wasserstein distance \cite{Gibbs2002} of the distributions.
The box plot gives the median as a black line, the 25\% to 75\% quartile as a yellow box and minimum and maximum value as the whiskers.
\label{fig:ViolinPlot_AlphaVentus}
}
\end{figure}

\subsection*{Superstatistics}
\label{sec:superstatistics}

Wind persistence statistics do not follow exponential distributions but require a refined statistical description. To appropriately describe the observed heavy tails, we consider $q$-exponentials as a generalization of exponential distributions \cite{Beck2001,Beck2003}. These generalized $q$-distributions have recently been used to describe waiting times in rainfall statistics \cite{Yalcin2016}, non-Gaussian diffusion processes \cite{Chechkin2017} or fluctuations in the frequency of the power grid \cite{Schaefer2018}.
$q$-exponentials are characterized by a $q$-parameter that determines the tails of the distribution and indicates heavy tails for $q>1$. In addition, a shape parameter $\lambda_q$ gives the decay rate so that the probability density reads \cite{Tsallis2009a}
\begin{equation}
   p(d|\lambda_q)=(2-q)\lambda_q\left[1-(1-q)\lambda_q d\right]^{\frac{1}{1-q}},
\label{eq:Q_PDF}
\end{equation}
which becomes an exponential in the limit $q\rightarrow 1$.
The kurtosis of $q$-exponential distributions is given as 
\begin{equation}
\kappa_{q \text{-exp}}=\frac{9}{5}+\frac{81}{30-25q}+\frac{1}{q-2}+\frac{8}{4q-5}
\label{eq:QKurtosis} 
\end{equation}
 and allows for arbitrarily large values, as it diverges at $q=\frac{6}{5}=1.2$ \cite{Anvari2016}, see also Supplementary Note 4 for an illustration.

Analyzing the persistence statistics, we indeed observe that $q$-exponentials are a better fit to the data than exponentials, see Fig.~\ref{fig:ExpAndQExpFits_lowWind} for low-wind  states of several locations in Europe. High-wind states display similar statistics (Supplementary Note 2) and $q$-exponentials are a better fit to the data than exponentials, based on likelihood analysis (Supplementary Note 6).
An important property of $q$-exponentials is that for $q>1$ and large arguments, the distribution follows a power law with exponent $1/(1-q)$:
\begin{equation}
   p(d)\sim d^{\frac{1}{1-q}},\text{ as } d\rightarrow \infty.
\label{eq:QPowerLaw}
\end{equation}
Hence, in particular the tails, i.e., the essential extreme event statistics, are well-captured using $q$-exponentials.

Next, we exploit that $q$-exponentials are not an arbitrary distribution with heavy tails but allow a deeper insight into the system's statistical properties, using \emph{superstatistics} \cite{Beck2001,Beck2003}. 

Suppose our data consists of samples drawn from different exponential distributions with different decay constants $\lambda_e$. If the decay constants  $\lambda_e$ are distributed following a $\chi^2$-distribution $g(\lambda_e)$, then the integral 
\begin{equation}
   p(d)= \int_0^\infty g(\lambda_e) p(d|\lambda_e) \text{d}\lambda_e
  \label{eq:integrate_exponentials}
\end{equation}
yields a $q$-exponential distribution (\ref{eq:Q_PDF}). That is, superimposing multiple exponentials leads to $q$-exponentials, if the constants $\lambda_e$ are distributed accordingly. 
Notably, the exact distribution of the decay constants $\lambda_e$ is of minor importance for $q$-values close to one. 
Hence, the $q$-exponential estimates reported in Fig.~\ref{fig:ExpAndQExpFits_lowWind} arise generically for any sharply peaked distribution $g(\lambda_e)$ \cite{Beck2001,Beck2003}. 

Can the observed $q$-exponentials in the wind persistence be explained using superstatistics? The wind data was recorded under very different atmospheric conditions, for example certain parts were recorded during a strong western circulation, while other periods were recorded during a large scale blocking situation
over Europe. To understand the observed persistence statistics as a superposition of individual distributions, we 
disaggregate the data into chunks with approximately homogeneous atmospheric conditions. In particular, we classify the large-scale atmospheric conditions according to a circulation weather type (CWT) approach \cite{Jones1993}.
CWTs describe the characteristics of the near-surface flow in terms of direction and intensity based on mean sea level pressure (MSLP) field around a reference point \cite{Jones1993}. For our study we use MSLP data from ERA-Interim and the reference point is located in Central Europe at 10\textdegree East and 50\textdegree North (near Frankfurt/Main). For this domain, the CWT directions are classified either as one of the eight cardinal and intercardinal directions (North, North-East, East, ...)  or a cyclonic/anti-cyclonic CWT, neglecting mixtures of cyclonic and directional CWTs. The strength of the flow is quantified using the $f$-parameter, which estimates the gradient of the instantaneous MSLP field around the reference point, and can thus be used as a proxy for the large-scale geostrophic wind (see Methods for details). Typical values for the $f$-parameter for Central Europe are 5 to 50 hPa/1000km, see \cite{Reyers2015} and Methods. Notably, assigning instantaneous weather types via $f$-parameters and CWT directions, allows a dynamical description of the synoptic state.
Using this approach, we decompose the data based on the dominant CWT direction or the $f$-parameter. Alternatively, we simply use the different recording years.

Indeed, disaggregating the data into small chunks of coherent $f$-parameters, leads to a lower kurtosis in the individual chunks and therefore better approximations by exponentials, see Fig.~\ref{fig:IndividualExpFits} (a)-(c). Hence, for a given $f$-parameter, the waiting process is better approximated by a Poisson process than it was when using measurements from the full period of interest. 
We also quantify this statement further by comparing the result to an alternative decomposition based on the recording year and to a plain Poisson distribution (Fig.~\ref{fig:ViolinPlot_AlphaVentus}). The distribution conditioned on the $f$-parameter has a smaller Wasserstein distance \cite{Gibbs2002} to any of the
1000 randomly drawn Poissonian realizations than the distribution conditioned on the recording year.
Furthermore, disaggregating the data according to the flow direction of the CWTs instead of the $f$-parameters does not reproduce the $q$-exponential equally well, see Supplementary Note 4. These surprising results will be examined in more detail in the synoptic section below.

As a consistency check of the superstatistical approach, we explicitly carry out the superposition of the individual exponential distributions found for different $f$-parameters, see Fig.~\ref{fig:IndividualExpFits} (d). 
Superposition and $q$-exponential agree very well for the persistence of high-wind situations at the off-shore wind farm Alpha Ventus. The agreement is not as good for low-wind situations at Harthaeuser Wald, where the superposition only partly explains the shape of the distribution. Finally, the $\lambda_e$ distribution approximately follows a log-normal distribution, 
a commonly observed distribution in superstatistics \cite{Beck2001,Beck2003}, see Fig.~\ref{fig:IndividualExpFits} (e).

We conclude that the heavy-tailedness of the full persistence statistics is at least partly explained as a result of the superposition process. Hence, the results support the idea of $q$-exponentials arising from a superposition of different conditioned distributions and highlight the importance of large-scale atmospheric conditions. 
Next, we investigate whether the persistence of aggregated wind power generation can also be described in terms of $q$-exponentials.

\subsection*{Power generation}
\label{sec:powergen}

\begin{figure*}
\begin{centering}
\includegraphics[width=1.9\columnwidth]{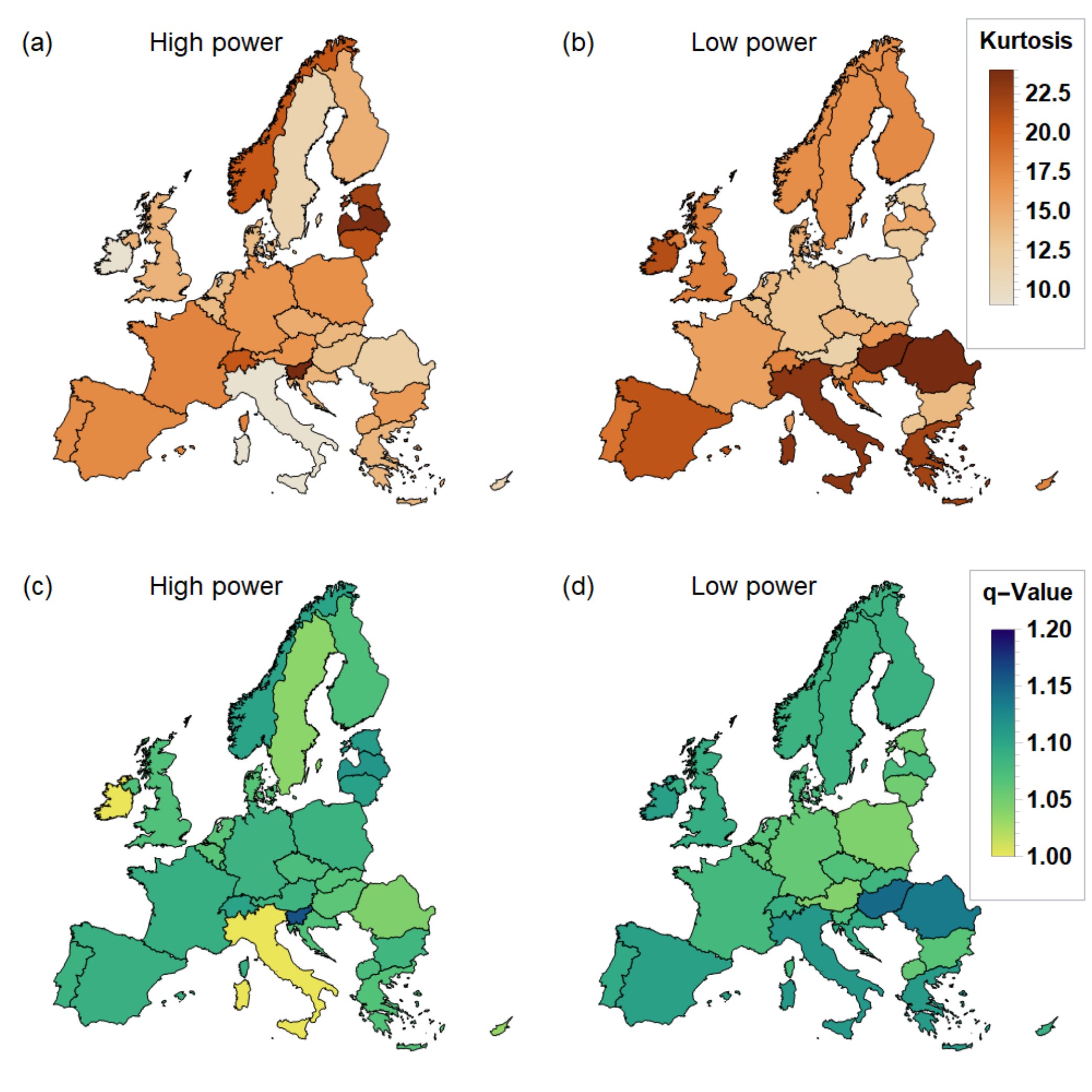}
\par\end{centering}
\caption{\textbf{European wind power generation persistence statistics are heavy-tailed.} European maps are shown with wind power generation data aggregated per country, based on the renewables.ninja data \cite{Pfenninger}.  An output is classified as high in panels~(a) and (c) if it is above the 75th quantile and as low in panels~(b) and (d) if it is below the 25th quantile. Panels~(a) and (b) give the kurtosis of the power persistence statistics. In addition, the $q$-parameter is computed based on eq. \eqref{eq:QKurtosis} and displayed in panels~(c) and (d). A high kurtosis consequently implies large $q$-values. Dark colors indicate a high kurtosis or $q$-value respectively. Note that $q=1.2$ is the maximum $q$-value, while we only plot the kurtosis up to 24.
Heavy tails and high $q$-parameters are especially prevalent for low power output and around the Mediterranean.
This analysis used aggregated data of off- and onshore wind generation per country. Distinguishing does not change the heavy tails significantly, see Supplementary Note 3.  
Maps were created using Wolfram Mathematica 11: \protect\url{https://www.wolfram.com/mathematica/}.
}
\label{fig:European_maps}
\end{figure*}

Not only the wind velocities, but also wind power generation time series exhibit extremely long periods of persistent low or high values. 
To show this, we analyze aggregated wind power generation time series documented in the renewables.ninja dataset v.1.1 obtained for the period 1980-2016 \cite{Pfenninger}, see Fig.~\ref{fig:European_maps}.
This analysis has three benefits: It directly discusses wind power instead of wind velocity, which have an approximately fixed relationship $P \sim v^3$ \cite{Ackermann2005}. Thereby, our statistical analysis becomes more applicable to the energy sector. Secondly, we consider wind power generation of whole countries instead of single locations and therefore refer to the importance of high- and low-wind power output in entire power systems. Furthermore, we verify early results by using the independent renewables.ninja dataset.

As before, the duration $d$ of periods, where the generation is constantly lower or higher than a reference value, is recorded. Specifically, for each country all generation above the 75th quantile is classified as high power and generation below the 25th quantile as low power. Again, we observe heavy-tailed distributions, i.e., a kurtosis higher than the expected value of $\kappa(\text{exp})=9$ both for periods of high power generation (panels~(a) and (c)), as well as for low power generation (panels~(b) and (d)). These observations are connected to the superstatistical approach by computing the $q$-parameter using eq. \eqref{eq:QKurtosis} and solving for $q$.

The Balkan, the Mediterranean, UK and Scandinavia show particularly heavy tails for low power generation, leading to the highest $q$-values, i.e., the most pronounced power laws in low-wind generation persistence statistics. Therefore, periods without wind generation have to be expected to last longer than based on a simple Poissonian statistics. Interestingly, Italy and Ireland display no heavy tails for high-power generation.
We finally note that the observed $q$-values are very similar to the ones recorded for the wind velocity persistence statistics based on the downscaled ERA-Interim data set (comparing  Fig.~\ref{fig:ExpAndQExpFits_lowWind} and Fig.~\ref{fig:European_maps}). 

Concluding, we also observe heavy tails in the wind power generation on a country-scale. Therefore, extreme events such as long periods with low wind speeds have to be considered when dimensioning energy storage and storage needs are likely to be higher than based on simple exponential estimates, see also Supplementary Note 8. We proceed with a synoptic view on these long waiting times.

\subsection*{Synoptic analysis}
\label{sec:synoptic}

\begin{figure*}
\begin{centering}
\includegraphics[width=1.9\columnwidth]{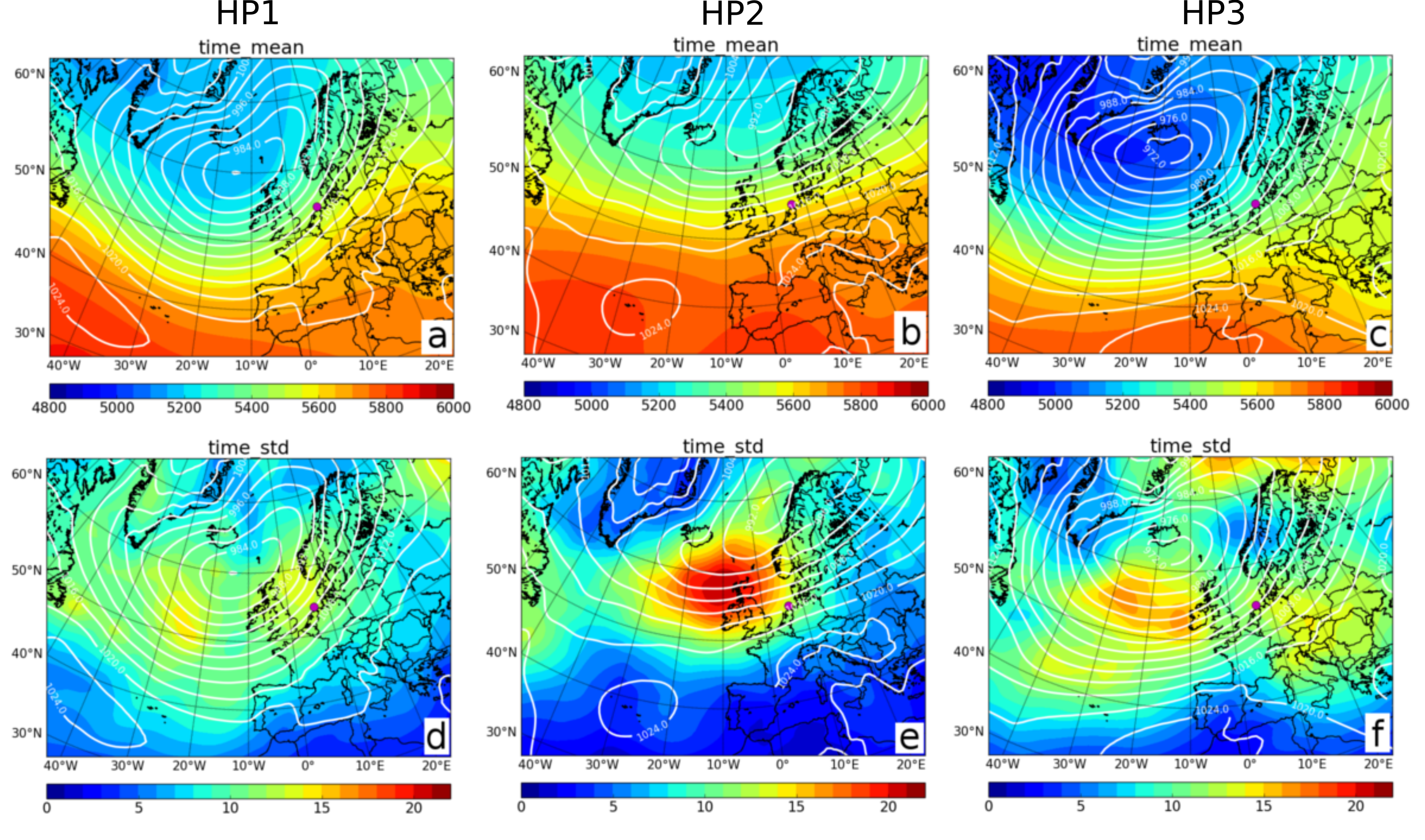}
\par\end{centering}
\caption{\textbf{High-wind periods are associated with different CWTs.}
The three columns illustrate the large scale atmospheric conditions as obtained by ERA-Interim during three extremely long high-wind periods: (a,d) HP1 in November-December 2006, (b,e)
HP2 in October 1983 and (c,f) HP3 in  January-February 1990.
The contours show the average mean sea level pressure (MSLP) in hPa while the shading shows the 
$500$~hPa geopotential height in meters (upper row, a-c), and the standard deviation of MSLP (lower row, d-e), respectively. The magenta dot shows the location of Alpha Ventus. 
 Maps were created using Python 2.7.12: \protect\url{https://www.python.org/}.
}
\label{fig:SynAlpha}
\end{figure*}

\begin{figure*}
\begin{centering}
\includegraphics[width=1.9\columnwidth]{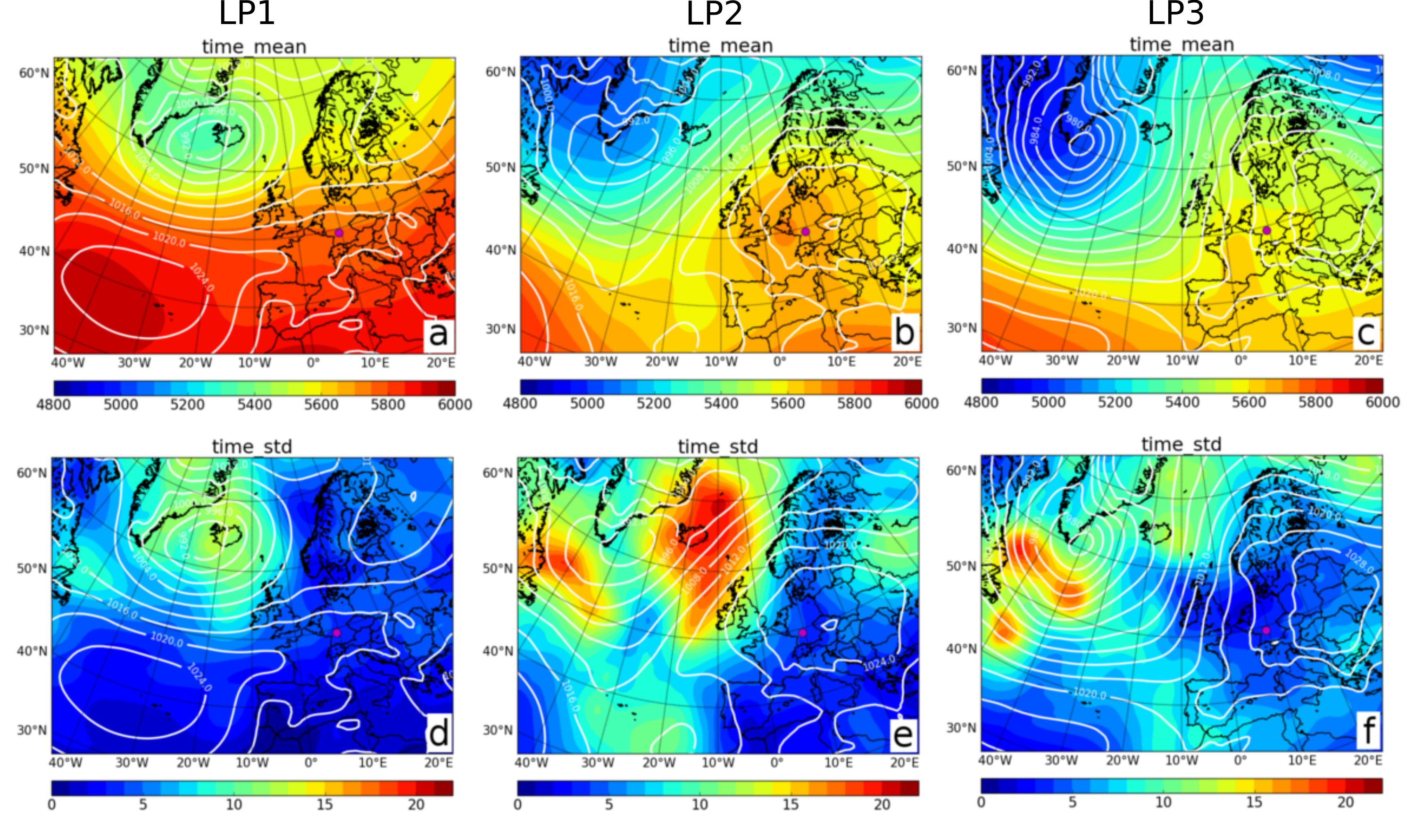}
\par\end{centering}
\caption{\textbf{Low-wind periods are associated with different CWTs.}
The three columns illustrate the large scale atmospheric conditions as obtained by ERA-Interim during three extremely long low-wind periods: (a,d) LP1 in August 2008, (b,e)
LP2 in November-December 1991 and (c,f) LP3 in December 1989-January 1990.
The contours show the average MSLP in hPa while the shading shows the 
500hPa geopotential height in meter (upper row, a-c), and the standard deviation of MSLP (lower row, d-e), respectively. The magenta dot shows the location of Harthaeuser Wald.    
 Maps were created using Python 2.7.12: \protect\url{https://www.python.org/}.
}
\label{fig:SynHarth}
\end{figure*}

Disaggregating the data using the $f$-parameters, a proxy for the pressure gradient (Fig. \ref{fig:IndividualExpFits}), approximates the superstatistical $q$-exponentials, while a separation of the data according to the flow direction of the CWTs does not reproduce the shape of the persistence distributions (Supplementary Note 4). This can be attributed to the intermittency of the atmospheric flow \cite{Stull1988,Nakamura2005} and indicates that both, prolonged calms and strong-wind situations
do occur for different and non-stationary CWTs in contrast to the naive expectation that high-wind periods occur solely for westerly CWTs and low-wind-periods solely for anti-cyclonic CWTs. Furthermore, high-wind periods may be distinctly longer than low-wind periods, based on the analysis of duration distributions shown in 
Fig.~\ref{fig:ExpAndQExpFits_lowWind} and Supplementary Note 2. This seems unexpected, as surface cyclones, which are associated with high wind speeds, usually pass Europe within only a few days due to their typical propagation velocity \cite{Loeptien2008}, while atmospheric blocking events, which may cause long-lasting calms, can have a lifetime of up to several weeks \cite{Brunner2017}. We perform a synoptic analysis of selected high- and low-wind periods to examine these findings.

Analyzing the downscaled ERA-Interim data set, all high-wind periods at Alpha Ventus that persist for more than 100 hours occur in the winter half year (October to March), except from one event (September 2004). On the other hand, low-wind periods at Harthaeuser Wald of 100 hours and more arise throughout the year.
Hence, for the analysis we simply select the three longest periods of the high-wind situations at Alpha Ventus (referred to as HP1, HP2, and HP3), while for Harthaeuser Wald we choose the two most persistent low-wind situations (both in winter) and the longest period that occurred in summer (LP1, LP2, and LP3). The precise periods are noted in the Methods.

HP1 and HP3 display similar synoptic patterns, characterized by a pronounced mid-tropospheric trough over the North Atlantic and strong mean sea level pressure gradients over Western Europe (see Fig.~\ref{fig:SynAlpha}). Accordingly, both periods are dominated by south-westerly CWTs (including some mixed classes). Snapshots at instantaneous points in time of HP1  reveal that a recurrent trough over the North Atlantic is existent throughout this period, though its amplitude varies (see Supplementary Note 5). Due to the trough, various strong and quasi-stationary surface steering cyclones develop between Iceland and the UK, with core pressures of partly below $960$~hPa. Hence, Alpha Ventus is continuously at the foreside of a rotating low pressure field (Fig.~\ref{fig:SynAlpha}a and \ref{fig:SynAlpha}c), which is reflected by the moderate standard deviations in the MSLP fields over Western Europe and the UK (Fig.~\ref{fig:SynAlpha}d and \ref{fig:SynAlpha}f). Snapshots for HP3 show similar pressure patterns (not shown). A different picture is revealed for HP2, which is characterized by a zonal mid-tropospheric flow, with Alpha Ventus being located at the southern flank of a stretched band of low surface pressure, with extended high pressure further south (Fig.~\ref{fig:SynAlpha}b). The high standard deviation in the MSLP field near the UK (Fig.~\ref{fig:SynAlpha}e) suggests that several synoptic systems (i.e. primarily lows) pass over the British Isles towards Northern Europe within the period. Snapshots indicate that the cyclones rapidly migrate along a north-easterly track towards Scandinavia (Supplementary Note 5). As a result, Alpha Ventus is permanently in the sphere of influence of alternating surface lows and highs during HP2, whereat the pressure gradients remain strong. In this case, nearly half of the CWTs detected in HP2 are anti-cyclonic, otherwise westerly but also northerly CWTs occur. A common characteristic of the three high-wind periods analyzed here is the clustering of strong surface cyclones \cite{Pinto2014}.

Low-wind situations, i.e., long calms, are similarly associated with predominantly but not exclusively anticyclone weather types, see Fig.~\ref{fig:SynHarth}.  For example, the summer event LP1 (Fig.~\ref{fig:SynHarth}a) exhibits a strong Azores High and extended ridge towards Central Europe. Accordingly, pressure gradients are weak at Harthaeuser Wald. A standard deviation of nearly zero suggests that the high pressure conditions are very stable and persist for the majority of the period (Fig.~\ref{fig:SynHarth}d). An atmospheric blocking over Central Europe is present during LP2 (Fig.~\ref{fig:SynHarth}b). The associated stable surface high exhibits very weak gradients near Harthaeuser Wald. The standard deviation of the MSLP field is low (Fig.~\ref{fig:SynHarth}e) and aside from anti-cyclonic mainly mixed anti-cyclonic/southerly CWTs occur in the period. During LP3, cold upper-level air lies over Eastern Europe (Fig.~\ref{fig:SynHarth}c). Below, a cold high pressure centre forms at the surface, which persists for several days. As Harthaeuser Wald is located at the western flank of the cold high, LP3 is dominated by southerly and anti-cyclonic CWTs. Again, pressure gradients and the standard deviation are low at Harthaeuser Wald (\ref{fig:SynHarth}f), hence conditions for a long low-wind period are fulfilled.

In summary, situations with persistent high or low wind speed conditions are not necessarily linked to recurring individual CWTs, which might be an explanation for our finding that superstatistics regarding the direction of the CWTs is not straightforward and provides mixed results.

\section*{Discussion}
In a fully renewable power system, the operation of storage, backup and sector coupling technologies will be crucially determined by periods of both low and high power feed-in by renewable generators \cite{Kempton2005,Paraschiv2014,Elsner2016}.
Long, persistent periods of extreme wind output are especially problematic, as most scenarios of highly renewable power systems use a high share of wind energy \cite{Heide2010,Rodriguez2015}. Complementary, long periods with high wind velocities determine how large back-up battery options or Power-to-Gas storage have to be dimensioned to not waste wind electricity \cite{Sternberg2015}.  Here, we have analyzed the persistence (waiting time) statistics of wind power, highlighting several interesting statistical observations.

Persistence statistics of wind velocities and wind power generation do not follow exponential distributions as intuitively expected \cite{Ross2014}, but display heavy tails (Fig.~\ref{fig:Era-InterimKurtosis}). Therefore, long periods of high-wind power output and periods of low-wind power output occur more often than based on simple Poissonian statistics.

While not perfect, a better description of the wind persistence statistics is found in  $q$-exponentials (Fig.~3), which are based on superstatistics, enjoying recent attention in time series analysis \cite{Yalcin2016,Chechkin2017,Schaefer2018}. 
We have revealed a superposition of several, atmospheric conditions as a potential mechanism giving rise to $q$-exponentials, in particularly when conditioning with respect to the $f$-parameter. The so derived $q$-exponentials allow a deeper insight into the underlying local dynamics than for example stretched exponentials would \cite{bunde2003effect,bunde2005long}.
Modeling wind persistence statistics as  $q$-exponentials does not only provide a good fit but also reveals a scaling law for the heavy tails, based on the $q$-value, going beyond previous investigations \cite{Simiu1996}.  Furthermore, our findings imply that the extreme event statistics of wind is governed by Fr\'echet distributions, instead of Gumbel statistics \cite{Tsallis2009a}, altering risk estimates, see e.g. \cite{Rabassa2014} for a detailed discussion. \revise{This is particularly remarkable as our finding could change an often used paradigm of extreme value statistics in wind engineering \cite{harris2009ximis}.} Energy storage capacities grow substantially when including the observed heavy tails in the analysis (Supplementary Note 8). Hence, future research on storage dimensioning should include our statistical findings to save costs due to failures caused by too small back-up systems.

Not only wind velocity persistence statistics are heavy-tailed but also wind power generation  persistence statistics are.  In particular, the duration of periods with low-wind power generation displays heavy tails. This demonstrates that our analysis is robustly applicable to countries as well as to individual locations and to different data sets. Using European (Fig.~\ref{fig:European_maps}) and in the future global data allows us to identify regions with particularly high risk of extremely long waiting times.

Our results are based on the well-established ERA-Interim reanalysis dataset \cite{Dee2011,Tobin2016,Moemken2018}, downscaled using the established RC4 regional model. Alternative regional models are expected to yield the same results, based on previous comparisons of regional models \cite{Hueging2013,Moemken2018}.

A synoptic analysis revealed that long low-wind periods are typically associated with very stable synoptic patterns such as blocking but also atmospheric ridges. In contrast, the synoptic conditions can be much more dynamic during high-wind periods, i.e., the instantaneous weather type changes over time. A clustering of surface cyclones led to high-wind periods lasting longer than three weeks, thus more persistent than the observed low-wind periods.
The direction of the large scale geostrophic wind changed during both low- and high-wind situations, particularly for the latter, such that considering the persistence of a single CWT in terms of its direction is not a suitable predictor for the duration of a low or high-wind period.

Concluding, we emphasize the role of persistence (waiting time) statistics when analyzing wind statistics, in particular based on its role for future energy systems. The presented superstatistical approach offers a new perspective on how to analyze wind data and a coherent framework to understand wind persistence statistics as a superposition of homogeneous wind and weather conditions. In particular, scaling of heavy tails and extreme event statistics are quantitatively determined by $q$-exponential distributions, which should be helpful for forecasts of extreme weather events or in dimensioning backup options in future energy systems, complementing existing analysis \cite{Steinke2013,Luo2015}.
We also complement the observations of $q$-exponential distributions of wind turbulence \cite{Manshour2016} by investigating spatially large scale systems (European continent) and longer time scales. 

However, many open questions remain. The emergence of heavy tails was modelled by using superpositioned $f$-parameters. If the time series of a given location were sufficiently long, the data could be split both for homogeneous $f$-parameter and individual CWT directions. The analysis could also be re-done for groups of related CWTs. 
Furthermore, frequency and persistence of CWTs may be affected by climate change, which is projected to change the temporal statistics of wind power generation \cite{Reyers2015, Weber2017, Wohland2017}.
While the current analysis focused on Europe, future work should consider other regions in the mid-latitudes to observe global scale atmospheric patterns influencing wind velocities and local CWTs. 
\revise{Furthermore, the extreme value statistics studied here could also be applied to waiting times of extreme wind gusts on longer time periods.}

Finally, our results already show that not only average wind velocities or their increments but additional meteorologic information, such as dynamically changing CWT directions and $f$-parameters, have to be included when analyzing wind statistics, and should be used in energy system analysis and design \cite{Grams2017}.

\section*{Material \& Methods}

\subsection*{Computing wind speed at turbines}

The downscaled ERA-Interim data provides a fine grid over Europe with wind speeds at 10 meters above the ground.
Since the hub height of wind turbines is typically around 100 meters above ground  \cite{Tobin2016}, the near-surface  wind velocities have to be extrapolated to a higher altitude. Assuming wind velocities increase algebraically with height \cite{Manwell2010}, we use the following power law formula of the wind speed $v(z)$ at height $z$: 
\begin{equation}
   v(z) = v_{z_0} \, (z/z_0)^{1/7},
\label{eq:Wind extrapolation}
\end{equation}
with $z_0 = 10~{\rm m}$ \cite{Dee2011} and $z = 100~{\rm m}$. 

\subsection*{Kurtosis and fitting $q$-exp}

Given $M$ measurements of the quantity $x_i$ with $\mu$ and $\sigma$ being the mean and standard deviation of the distribution, respectively, the kurtosis is given as the normalized 4th moment by 
\begin{equation}
   \kappa := \frac{1}{M}\sum_{i=1}^{M}\left(\frac{x_{i}-\mu}{\sigma}\right)^{4}.
\end{equation}
\revise{Contrary to some notations of the kurtosis as "peakedness", the kurtosis should be seen as a measure for the heavy tails of a distribution \cite{westfall2014kurtosis}.} Some studies may apply the excess kurtosis, which is obtained by subtracting the kurtosis of a Gaussian distribution $\kappa_\text{Gauss}=3$,
\begin{equation}
   \kappa_\text{Excess}=\kappa-3.
\end{equation}

For the exponential distribution, the kurtosis is given as
\begin{equation}
\kappa_\text{exp}=9.
\end{equation}
A kurtosis higher than nine, $\kappa>9$, points to heavy tails, i.e., increased likelihood of very large values.


To re-iterate, the kurtosis of $q$-exponentials is given by 
\begin{equation}
\kappa_{q \text{-exp}}=\frac{9}{5}+\frac{81}{30-25q}+\frac{1}{q-2}+\frac{8}{4q-5}.
\label{eq:QKurtosis2} 
\end{equation}

Searching for an adequate description of the recorded wind persistence statistics, we compute the best-fitting $q$-exponential distribution as follows. First, we compute the kurtosis and then determine the value of the resulting $q$ via Eq.~\eqref{eq:QKurtosis2}. Next, we perform a maximum likelihood analysis to find the most likely value for $\lambda_q$. This ensures that especially the tails of the distributions are fitted accordingly, as the $q$-parameter determines the power-law scaling of the $q$-exponential.

\subsection*{Assigning weather conditions}

To determine the circulation weather type (CWT), ERA-Interim data \cite{Dee2011} of the atmospheric conditions over Europe were considered. Specifically, instantaneous daily mean sea level pressure (MSLP) fields around a reference point in Central Europe (10\textdegree East and 50\textdegree North near Frankfurt/Main, Germany) were used. The CWT classes consist of eight directional weather types (e.g. 'North', 'South-West', 'West', etc.) and two rotational weather types ('Cyclonic' or 'Anti-cyclonic'), depending on the dominant part of the flow. Of special interest for the current analysis is the $f$-parameter, which estimates the gradient of the instantaneous MSLP field at the reference point:
\begin{equation}
   f = \sqrt{\left(\frac{\partial p}{\partial x}\right)^2 + \left(\frac{\partial p}{\partial y}\right)^2},
\label{eq:f-parameter-definition}
\end{equation}
with $\frac{\partial p}{\partial x}$ and $\frac{\partial p}{\partial y}$ being the zonal and meridional pressure gradients, respectively. This parameter can thus be used as a proxy for the large-scale geostrophic wind: Large $f$-parameters indicate higher pressure gradients, and thus typically higher wind speeds, see \cite{Reyers2015,Weber2017,Donat2010} for details.

While the downscaled ERA-Interim data set uses a 3-hour resolution for the wind speed \cite{Jacob2014}, the available weather data \cite{Reyers2015} 
assigns one $f$-parameter and one CWT per day. Hence, we assume the weather type and $f$-parameter to be identical for all 3-hour intervals during one day, when comparing with the wind speed.

Typically, low- and high-wind episodes endure several days and may feature more than one (typically related) CWTs. This means that a high-wind situation can include multiple days with potentially different CWT or $f$-parameters. In these cases, we use the dominant CWT (using the first occurring one in cases of ties) and compute the average $f$-parameter of the period.

\subsection*{Superstatistics}

The following formula illustrates how a superposition of
ordinary exponentials, given a $\chi^2$-distribution of the exponents, leads to a $q$-exponential:

\begin{displaymath}
\int_0^\infty d \beta f(\beta) e^{-\beta E}
=\frac{1}{(1+(q-1)\beta_0 E)^{1/(q-1)}},
\end{displaymath}
where
\small{
\begin{displaymath}
f (\beta) = \frac{1}{\Gamma \left( \frac{1}{q-1} \right)} \left\{
\frac{1}{(q-1)\beta_0}\right\}^{\frac{1}{q-1}}
\beta^{\frac{1}{q-1}-1} \exp\left\{-\frac{\beta}{(q-1)\beta_0}
\right\} 
\end{displaymath}}
is the $\chi^2$-distribution, with Gamma function $\Gamma$.
In the general statistical mechanics formalism, $E$ is the energy and $\beta$ a fluctuating
inverse temperature parameter \cite{Beck2001,Beck2003,Tsallis2009a}. For our application to persistence statistics, we
identify $E=d$ and $\beta =\lambda_e$.

\subsection*{Artificial Poissonian}
Let us explain the procedure leading to Fig.~\ref{fig:ViolinPlot_AlphaVentus} in more detail. The downscaled ERA-Interim data at Alpha Ventus for the 31 years consists of 90584 velocity measurements. To generate artificial data, we first approximate the persistence statistics of these wind data with an exponential distribution, see Fig.~\ref{fig:Era-InterimKurtosis}. Then, we simulate a Poisson process with a rate given as the estimated exponential decay rate and a total of 90584 data points are generated to have an equal number of artificial and real "measurements".
Next, the real data is split into 31 evenly sized data packages. First, the separation is done based on that each package has an approximately homogeneous $f$-parameter. Since some $f$-parameters are more likely to occur, the intervals of the $f$-parameters are not homogeneous. As an alternative, we split the data based on the year of recording the data.
Finally, the artificial Poissonian data is also split into 31 packages. For all packages, we compute the persistence statistics and the kurtosis and $q$-value thereof.

\subsection*{Uncertainties of parameters}
To estimate the uncertainty of our stochastic estimates, we make use of \emph{bootstrapping} \cite{Efron1994,Bohm2010}: Given a number of measurements $(x_1, x_2, ..., x_{N_m})$, in our case duration values, we can compute stochastic quantities such as the kurtosis or perform exponential fits. 
Instead of doing this only once for the full data set using each values only once, we draw randomly $N_m$ entries from our measurements, allowing for duplicates. With this new set of measurements $(\tilde{x}_1,\tilde{x}_2,...,\tilde{x}_{N_m})$ we again compute the kurtosis, find the best exponential fit etc. This procedure is repeated $N_b$ times so that we obtain a mean kurtosis and a mean exponential fit but also a standard deviation of the kurtosis estimate and so on. The uncertainties of the $q$-values are all included explicitly in the figures. 
Overall, the relative errors are of the following order:  $\Delta {\lambda}_E \sim 1-2\%$, $\Delta {\lambda}_q \sim 2-8\%$, $\Delta {\kappa} \approx \Delta q \sim 2-5\%$.

\subsection*{Selecting persistent events for synoptic analysis}
When performing the synoptic analysis, we chose the following high pressure (HP) and low pressure (LP) events: For Alpha Ventus the periods are 13 November to 08 December 2006 (609 hours: HP1), 10 October to 28 October 1983 (435 hours; HP2), and 29 January to 16 February 1990 (432 hours; HP3). For Harthaeuser Wald we selected the periods 21 August to 30 August 2008 (210 hours; LP1), 25 November to 05 December 1991 (255 hours; LP2), and 29 December 1989 to 07 January 1990 (237 hours; LP3).


\begin{acknowledgments}
\textbf{Funding:} We gratefully acknowledge support from the Federal Ministry of Education
and Research (BMBF grant no. 03SF0472A-F and 03EK3055F), the Helmholtz
Association (via the joint initiative ``Energy System 2050 - A Contribution
of the Research Field Energy'' and the grant no.VH-NG-1025), the EPSRC (grant EP/N013492/1) and the German Science Foundation (DFG) by a grant toward the Cluster of Excellence ``Center for Advancing Electronics Dresden'' (cfaed). J.G.P. thanks the AXA Research Fund for support. This project has received funding from the European Union’s Horizon 2020 research and innovation programme under the Marie Sk\l{}odowska--Curie grant agreement No 840825.
\textbf{Author contributions:}  J.W., D.W. and B.S. conceived and designed the research. J.W., M.R. and B.S. acquired the data and performed the data analysis.  
C.B. and B.S. performed the superstatistical analysis. 
M.T., J.G.P. and all other authors contributed methods and analysis tools, discussed and interpreted the results and wrote the manuscript. 
\textbf{Competing interests:} The authors declare no competing interests.
\textbf{Data availability:} Raw data for the wind velocity analysis are available from the ESGF Node at DKRZ at \url{https://esgf-data.dkrz.de/projects/esgf-dkrz/}.
Wind power generation data are available at \url{https://www.renewables.ninja/}.
All data that support the results presented in
the figures of this study are available from the authors upon reasonable request.
\end{acknowledgments}

\bibliographystyle{naturemag}

\end{document}